\documentclass[11pt]{article} 
\usepackage{latexsym}
\usepackage{amssymb}
\usepackage{amsmath}
\usepackage{mathrsfs}
\usepackage{amsfonts}
\usepackage{amsthm}
\usepackage{epsfig}

\input{amssym.def}

\newcommand{\weakly}{\mbox{$ \;\stackrel{\cal D}{\longrightarrow}\; $}}

\newcommand{\IND}{{\mathbb I}}

\newcommand{\PR}{\mbox{\rm Pr}} 
 
\newcommand{\VAR}{\mbox{\rm Var}}

\newcommand{\iid}{\mbox{i.i.d.}\!}

\def\be{\begin{eqnarray}}
\def\ee{\end{eqnarray}}
\def\ben{\begin{eqnarray*}}
\def\een{\end{eqnarray*}}

\def\elabel#1{\label{e:#1}}

\def\sq{$\Box$}

\def\qed{\ifmmode\sq\else{\unskip\nobreak\hfil
\penalty50\hskip1em\null\nobreak\hfil\sq
\parfillskip=0pt\finalhyphendemerits=0\endgraf}\fi\par\medbreak}

\newsavebox{\junk}
\savebox{\junk}[1.6mm]{\hbox{$|\!|\!|$}}

\def\til={{\widetilde =}}

 \def\eq#1/{(\ref{#1})}

\def\eq#1/{(\ref{e:#1})}

\newcommand{\beqn}[1]{\notes{#1}%
\begin{eqnarray} \elabel{#1}}

\newcommand{\eeqn}{\end{eqnarray} }

\newcommand{\beq}[1]{\notes{#1}%
\begin{equation}\elabel{#1}}

\newcommand{\eeq}{\end{equation}} 

\def\bdes{\begin{description}}
\def\edes{\end{description}}

\def\notes#1{}

\def\parsec{\par\noindent}

\def\med{\medskip\parsec}

\def\basswidth{2.5in}
\def\gutter{0.5in}

\begin{document}

\begin{center}
{\LARGE {\bf Identifying Statistical Dependence in Genomic

\medskip
Sequences via Mutual Information Estimates\footnotemark[1]}}

\vspace{10mm}

\begin{tabular}{lll}
Hasan Metin Aktulga      & Ioannis  Kontoyiannis     & L. Alex Lyznik \\
Dept. of Computer Science& Department of Informatics & \\
Purdue University        & Athens Univ of Econ \& Business & Pioneer Hi-Breed International\\
W. Lafayette, IN 47907   & Patission 76, Athens 10434& Johnston, IA\\
U.S.A.                    & Greece                    & U.S.A.\\
{\tt haktulga@cs.purdue.edu}& {\tt yiannis@aueb.gr}& {\tt alex.lyznik@pioneer.com}\\

\vspace{5mm} \\

Lukasz Szpankowski     & Ananth Y. Grama           & Wojciech Szpankowski\\
Bioinformatics Program & Dept. of Computer Science & Dept. of Computer Science\\
UCSD                   & Purdue University         & Purdue University\\
San Diego, CA          & W. Lafayette, IN 47907    & W. Lafayette, IN 47907\\
U.S.A.                  & U.S.A.                     & U.S.A.\\
{\tt lszpanko@bioinf.ucsd.edu} & {\tt ayg@cs.purdue.edu} & {\tt spa@cs.purdue.edu}
\end{tabular}
\end{center}

\footnotetext[1]{This research was supported in part by the NSF
Grants CCR-0208709, CCF-0513636, and DMS-0503742, and the
NIH Grant R01 GM068959-01.}

\begin{abstract}        

Questions of understanding and quantifying the 
representation and amount of information in organisms 
have become a central part of biological research, as 
they potentially hold the key to fundamental advances.
In this paper, we demonstrate the use of 
information-theoretic tools for the task 
of identifying segments of biomolecules (DNA or RNA) 
that are statistically correlated. We develop 
a precise and reliable methodology, based on
the notion of mutual information,
for finding and extracting statistical
as well as structural dependencies.
A simple threshold function is defined,
and its use in quantifying the level of 
significance of dependencies between
biological segments is explored. These
tools are used in two specific applications.
First, for the identification of correlations 
between different parts of the maize zmSRp32 gene.
There, we find significant dependencies between
the 5' untranslated region in zmSRp32
and its alternatively spliced exons. 
This observation may indicate the presence 
of as-yet unknown alternative splicing 
mechanisms or structural scaffolds. 
Second, using data from the FBI's Combined 
DNA Index System (CODIS),
we demonstrate that our approach 
is particularly well suited for the 
problem of discovering short tandem repeats, 
an application of importance in genetic profiling.

\end{abstract}
\med

\newpage

\section{Introduction}
\label{intro}

Questions of quantification, representation, and 
description of the overall flow of information in biosystems 
are of central importance in the life sciences. In this paper, 
we develop statistical tools based on information-theoretic
ideas, and demonstrate their use in identifying informative 
parts in biomolecules.  Specifically, our goal is to 
detect statistically dependent segments of biosequences,
hoping to reveal potentially important biological 
phenomena. It is well-known \cite{hag06,segal06,steuer02} 
that various parts of biomolecules, such as DNA, RNA, and 
proteins, are significantly (statistically) correlated,
although formal measures and techniques for quantifying 
these correlations are topics of current investigation. 
The biological implications of these correlations are deep, 
and they themselves remain unresolved. For example, 
statistical dependencies between {\it exons} carrying protein coding 
sequences and noncoding {\it introns} may indicate the existence 
of as-yet unknown error correction mechanisms or structural 
scaffolds. Thus motivated, we propose to develop precise and 
reliable methodologies for quantifying and identifying
such dependencies, based on the information-theoretic
notion of {\it mutual information}.

Biomolecules store information in the form 
of monomer strings such as deoxyribonucleotides, ribonucleotides,
and amino acids. As a result of numerous genome and protein sequencing
efforts, vast amounts of sequence data is now available for
computational analysis. While basic tools such as BLAST
provide powerful computational engines for identification of conserved
sequence motifs, they are less suitable for detecting potential
hidden correlations without experimental precedence (higher-order
substitutions).

The application of analytic methods for finding regions
of statistical dependence through mutual information 
has been illustrated 
through a comparative analysis of the 5' untranslated regions 
of DNA coding sequences \cite{osada06}.
It has been known that eukaryotic translational initiation
requires the consensus sequence around the start codon defined as the
Kozak's motif \cite{kozak99}. By screening at least 500 sequences,
an unexpected correlation between positions -2 and -1 of the Kozak's
sequence was observed, thus implying a novel translational initiation
signal for eukaryotic genes. This pattern was discovered using mutual
information, and not detected by analyzing single-nucleotide
conservation. In other relevant work, neighbor-dependent 
substitution matrices were applied to estimate the average 
mutual information content of the core promoter regions from five
different organisms~\cite{reddy06a,reddy06b}. 
Such comparative analyses
verified the importance of TATA-boxes and transcriptional initiation.
A similar methodology elucidated patterns of sequence conservation
at the 3' untranslated regions of orthologous genes 
from human, mouse,
and rat genomes \cite{shabalina04}, making them potential targets for
experimental verification of hidden functional signals.  

In a different kind of application,
statistical dependence techniques find important applications
in the analysis of gene expression data. 
Typically, the basic underlying assumption in such analyses
is that genes expressed similarly under divergent conditions
share functional domains of biological activity. 
Establishing dependency
or potential relationships between sets of genes from 
their expression profiles holds the key to the 
identification of novel functional elements.
Statistical approaches to estimation of mutual information from gene
expression datasets have been investigated in \cite{steuer02}.

Protein engineering is another important area where
statistical dependency tools are utilized. Reliable 
predictions of protein
secondary structures based on long-range dependencies may enhance
functional characterizations of proteins \cite{baldi99}. 
Since secondary
structures are determined by both short- and long-range interactions
between single amino acids, 
the application of comparative statistical
tools based on consensus sequence algorithms or short amino acid
sequences centered on the prediction sites, is far from optimal. 
Analyses that incorporate 
mutual information estimates
may provide more accurate predictions.

In this work we focus on developing reliable and precise
information-theoretic methods for determining whether two biosequences
are likely to be statistically dependent.
Another motivating factor for this project, which
is more closely related to ideas from information theory,
is the question of determining whether
there are error correction mechanisms built into 
large molecules, as argued by Battail; see \cite{battail}
and the references therein. We choose to work with protein
coding exons and non-coding introns. While exons
are well conserved parts of DNA, introns have much greater
variability. 
They are dispersed on strings of biopolymers 
and still they have to be precisely identified in order 
to produce biologically relevant information. It seems that there 
is no external source of information but the structure of RNA 
molecules themselves to generate functional templates 
for protein synthesis. Determining potential mutual relationships 
between exons and introns may justify additional search for still 
unknown factors affecting RNA processing.

The complexity and importance of the RNA processing system 
is emphasized by the largely unexplained mechanisms
of alternative splicing, which
provide a source of substantial diversity in gene products.
The same sequence may be recognized as an exon or 
an intron, depending on a broader context of splicing reactions. 
The information that is required for the
selection of a particular segment of RNA molecules is very likely 
embedded into either exons or introns, or both. 
Again, it seems that the 
splicing outcome is determined by structural information carried by 
RNA molecules themselves, unless the fundamental dogma of biology 
(the unidirectional flow of information from DNA to proteins) is to 
be questioned.

Finally, the constant evolution of genomes introduces certain
polymorphisms, such as {\it tandem repeats}, which are an important
component of genetic profiling applications. 
We also study these forms of statistical
dependencies in biological sequences 
using mutual information.

In Section~\ref{theory} we develop some
theoretical background, and we derive 
a threshold function
for testing statistical significance.
This function admits a dual interpretation
either as the classical log-likelihood
ratio from hypothesis testing,
or as the ``empirical mutual 
information.''

Section~\ref{experiments} contains
our experimental results:
In Section~\ref{DNASD} we present our empirical
findings on the problem of  
detecting statistical dependency between 
different parts in a DNA sequence.
Extensive numerical experiments were carried
out on certain regions of the maize 
zmSRp32 gene \cite{lyznik04},
which is functionally homologous to 
the human ASF/SF2 alternative 
splicing factor. The efficiency of the
empirical mutual information in this
context is demonstrated. Moreover,
our findings suggest the existence
of a biological connection between
the 5' untranslated region in zmSRp32
and its alternatively spliced exons.

Finally, in Section~\ref{STRs} we show
how the empirical mutual information
can be utilized in the difficult problem
of searching DNA sequences for short 
tandem repeats (STRs), an important task
in genetic profiling. We extend the
simple hypothesis test of the previous
sections to a methodology for testing
a DNA string against different ``probe''
sequences, in order to detect STRs
both accurately and efficiently.
Experimental results on DNA sequences 
from the FBI's Combined DNA Index System 
(CODIS) are presented, showing that 
the empirical mutual information 
can be a powerful tool in this context
as well.

\section{Theoretical Background}
\label{theory}

In this section we outline the theoretical basis for 
the mutual information estimators we will later apply
to biological sequences.

Suppose we have two strings of unequal lengths,
\begin{eqnarray}
X_1^n &=&
        X_1,X_2,\ldots,X_n\\
Y_1^M &=&
        Y_1,Y_2,Y_3,\ldots\ldots\ldots,Y_M\,,
\end{eqnarray}
where $M\geq n$, taking values in a common finite 
alphabet $A$.
In most of our experiments, $M$ is significantly
larger than $n$; typical values of interest are
$n\approx 80$ and $M\approx 300$.  Our main goal
is to determine whether or not there is some
form of statistical dependence between them.
Specifically, we assume that the string $X_1^n$
consists of independent and identically distributed
($\iid$) random variables $X_i$ with common distribution
$P(x)$ on $A,$ and that the random variables $Y_i$ are
also $\iid$ with a possibly different distribution $Q(y)$.
Let $\{W(y|x)\}$ be a family of conditional distributions,
or ``channel,'' with the property that, when the input
distribution is $P$, the output has distribution $Q$,
that is,
$\sum_{x\in A} P(x)W(y|x)=Q(y),\;\;\mbox{for all $y$.}$
We wish to differentiate between
the following two scenarios:

{\em (I) Independence. } $X_1^n$ and $Y_1^M$
        are independent.

{\em (II) Dependence. } First $X_1^n$ is generated.
        Then an index $J\in\{1,2,\ldots,M-n+1\}$ is chosen
        in an arbitrary way,
        and $Y_J^{J+n-1}$ is generated
        as the output of the discrete memoryless channel
        $W$ with input $X_1^n$; that is,
	for each $j=1,2,\ldots,n$,
        the conditional distribution of
        $Y_{j+J-1}$ 
        given $X_1^n$ is $W(y|X_j)$.
        Finally the rest of the $Y_i$'s are
        generated $\iid$ according to $Q$.
[To avoid the trivial case where
both scenarios are identical,
we assume
that the rows of $W$ are not all
equal to $Q$ so that in the second scenario
$X_1^n$ and $Y_J^{J+n-1}$ are actually {\em not} independent.]

It is important at this point to note that, although neither 
of these two cases is biologically realistic as a description 
of the elements in a genomic sequence, it turns out that this 
set of assumptions provides a good operational starting point:
The experimental results reported in Section~\ref{experiments}
clearly indicate that, in practice,
the resulting statistical methods obtained 
under the present assumptions can provide
accurate and biologically relevant information.

To distinguish between {\em (I)} and {\em (II)}, 
we look at every possible alignment of $X_1^n$ with $Y_1^M$,
and we estimate the mutual information between
them. Recall that for two random variables $X,Y$ with marginal 
distributions $P(x),Q(y)$, respectively, and joint distribution 
$V(x,y)$, the mutual information between $X$ and $Y$
is defined as,
\be
I(X;Y)=\sum_{x,y\in A} V(x,y)\log\frac{V(x,y)}{P(x)Q(y)}.
\label{eq:mi}
\ee
Recall also \cite{cover} that $I(X;Y)$ is always nonnegative,
and it equals zero if and only if $X$ and $Y$ are independent.
The logarithms above and throughout the paper are taken to base 2,
$\log=\log_2$, so that $I(X;Y)$ can be interpreted
as the number of bits of information that each of these two 
random variables carries about the other; cf.\ \cite{cover}.

In order to distinguish between the two scenarios above,
we compute the empirical mutual information
between $X_1^n$ and each contiguous substring of $Y_1^M$
of length $n$: For each $j=1,2,\ldots,M-n+1$, let
$\hat{p}_j(x,y)$  denote
the joint empirical distribution
of $(X_1^n,Y_j^{j+n-1})$, i.e.,
let $\hat{p}_j(x,y)$ be the proportion of
the $n$ positions in $(X_1,Y_j),(X_2,Y_{j+1}),\ldots,(X_n,Y_{j+n-1})$
where $(X_i,Y_{j+i-1})$ equals $(x,y)$. Similarly,
let $\hat{p}(x)$ and $\hat{q}_j(y)$
denote the empirical distributions of $X_1^n$ and
$Y_j^{j+n-1}$, respectively.
We define the empirical (per-symbol) mutual information
$\hat{I}_j(n)$ between $X_1^n$ and $Y_j^{j+n-1}$
by applying (\ref{eq:mi}) to the empirical instead
of the true distributions, so that,
\begin{equation}
  \hat{I}_j(n)=\sum_{x,y\in A}\hat{p}_j(x,y)\log
        \frac{\hat{p}_j(x,y)}{\hat{p}(x)\hat{q}_j(y)}.
  \label{eq:hat{I}_j}
\end{equation}
The law of large numbers implies that,
as $n\to\infty$, we have
$\hat{p}(x)\to P(x)$, $\hat{q}_j(y)\to Q(x)$,
and $\hat{p}_j(x,y)$ converges to the true joint 
distribution of $X,Y$. 

Clearly, this implies that 
in scenario {\em (I)}, where $X_1^n$ and $Y_1^n$
are independent,
$\hat{I}_j(n)\to 0$,
for any fixed $j$, as $n\to\infty$.
On the other hand in scenario {\em (II)},
$\hat{I}_J(n)$ converges to $I(X;Y)>0$ where the
two random variables $X,Y$ are such that $X$
has distribution $P$ and the conditional distribution
of $Y$ given $X=x$ is $W(y|x)$.

\subsection{An Independence Test Based on Mutual Information}

We propose to use the following simple {\em test}
for detecting dependence between $X_1^n$ and $Y_1^M$.
Choose and fix a threshold $\theta>0$,
and compute the empirical mutual information
$\hat{I}_j(n)$ between $X_1^n$ and each
contiguous substring $Y_j^{j+n-1}$
of length $n$ from $Y_1^M$.
If $\hat{I}_j(n)$
is larger than
$\theta$ for some $j$, declare that
the strings $X_1^n$ and $Y_{j}^{j+n-1}$ are
{\em dependent}; otherwise,
declare that they are {\em independent.}

Before examining the issue of selecting
the value of the threshold $\theta$,
we note that this statistic is identical to the
(normalized) log-likelihood ratio between the above two hypotheses.
To see this, observe that, expanding the definition
of $\hat{p}_j(x,y)$ in $\hat{I}_j(n)$,
we can simply rewrite,
\ben
\hat{I}_j(n) 
&=&
        \sum_{x,y\in A}
        \frac{1}{n}\sum_{i=1}^n\IND_{\{(X_i,Y_{j+i-1})\}}(x,y)
        \log\frac{\hat{p}_j(x,y)}{\hat{p}(x)\hat{q}_j(y)}\\
&=&
        \frac{1}{n}\sum_{i=1}^n
        \sum_{x,y\in A}
	\IND_{\{(X_i,Y_{j+i-1})\}}(x,y)
        \log\frac{\hat{p}_j(x,y)}{\hat{p}(x)\hat{q}_j(y)},
\een
where the indicator function
$\IND_{\{(X_i,Y_{j+i-1})\}}(x,y)$ equals 1 if $(X_i,Y_{j+i-1})=(x,y)$,
and it is equal to zero otherwise. Then,
\begin{eqnarray}
\hat{I}_j(n) 
  &=&
        \frac{1}{n}\sum_{i=1}^n\log\frac{\hat{p}_j(X_i,Y_{j+i-1})}
        {\hat{p}(X_i)\hat{q}_j(Y_{j+i-1})}
	\label{eq:alt}\\
  &=&
        \frac{1}{n}\log
	\left[
	\frac{\prod_{i=1}^n\hat{p}_j(X_i,Y_{j+i-1})}
        {\prod_{i=1}^n\hat{p}(X_i)\hat{q}_j(Y_{j+i-1})}
	\right],
	\nonumber
\end{eqnarray}
which is exactly the normalized logarithm of the 
ratio between the joint empirical likelihood
$\prod_{i=1}^n\hat{p}_j(X_i,Y_{j+i-1})$
of the two strings, and the 
product of their empirical marginal likelihoods
$[\prod_{i=1}^n \hat{p}(X_i)][\prod_{i=1}^n \hat{q}_j(Y_{j+i-1})]$.

\subsection{Probabilities of Error}

There are two kinds of errors this test can make: Declaring
that two strings are dependent when they are not, and vice versa.
The actual probabilities of these two types of errors depend
on the distribution of the statistic $\hat{I}_j(n)$.
Since this distribution
is independent of $j$, we take $j=1$ and write $I(n)$
for the normalized log-likelihood ratio $\hat{I}_1(n)$.
The next two subsections present some classical asymptotics
for $\hat{I}_1(n)$. 

\paragraph{Scenario (I): Independence. }
We already noted that in this case
$I(n)$ converges to zero as $n\to\infty$, 
and below we shall see that this
convergence takes place at a rate 
of approximately $1/n$. Specifically,
$I(n)\to 0$ with probability one,
and a standard application of the multivariate central
limit theorem for the joint empirical distribution $\hat{p}_j$
shows that $n I(n)$ converges in distribution to a (scaled)
$\chi^2$ random variable. This a classical result in
statistics \cite{lehmann,schervish}, and, in the present context, it
was rederived by Hagenauer et al.~\cite{hag04,hag05}.
We have,
$$(2\ln 2) n I(n)\weakly  Z\sim
\chi^2((|A|-1)^2),$$
where $Z$ has a $\chi^2$ distribution with
$k=(|A|-1)^2$ degrees of freedom, and where
$|A|$ denotes the size of the data alphabet.

Therefore, for a fixed threshold $\theta>0$ and large $n$,
we can estimate the probability of error as,
\begin{eqnarray}
P_{e,1}
  &=&
        \PR\{\mbox{declare dependence}\,|\,
        \mbox{independent strings}\}
	\nonumber\\
  &=&
        \PR\{I(n)>\theta\,|\,
        \mbox{independent strings}\}
	\nonumber\\
  &\approx&
        \PR\{Z>(2\ln 2)\theta n\},
	\label{eq:false_positives}
\end{eqnarray}
where $Z$ is as before. Therefore,
for large $n$ the 
error probability $P_{e,1}$ decays 
like the tail of the $\chi^2$ distribution function, 
$$P_{e,1}\approx 1-\frac{\gamma(k,(\theta\ln 2)n)}{\Gamma(k)},$$
where $k=\frac{(|A|-1)^2}{2}$, and
$\Gamma,\gamma$ denote the
Gamma function and the incomplete 
Gamma function, respectively.
Although this is fairly implicit,
we know that the tail of the $\chi^2$ distribution 
decays like $e^{-x/2}$ as $x\to\infty$, therefore, 
\begin{equation}
P_{e,1}\approx
\exp\Big\{-(\theta\ln 2) n\Big\},
\label{eq:p1}
\end{equation}
where this approximation is to first
order in the exponent.

\paragraph{Scenario (II): Dependence. }
In this case the
asymptotic behavior of
the test statistic $I(n)$ is somewhat different.
Suppose as before that the random variables $X_1^n$ are
i.i.d.\ with distribution $P$, and that the conditional
distribution of each $Y_i$ given $X_1^n$
is $W(y|X_i)$, for some fixed
family of conditional distributions $W(y|x)$;
this makes the random variables
$Y_1^n$ i.i.d.\ with distribution $Q$.

We mentioned in the last section that,
under the second scenario,
$I(n)$ converges to the true 
underlying value $I=I(X;Y)$ of the mutual 
information, but, as we show below,
the rate of this convergence
is slower than the $1/n$ rate of scenario {\em (I)}:
Here, $I(n)\to I$ with probability one,
but only at rate $1/\sqrt{n}$,
in that,
$\sqrt{n}[I(n)-I]$ converges in distribution to
a Gaussian,
\be
\sqrt{n}[I(n)-I]\weakly V\sim N(0,\sigma^2),
\label{eq:clt}
\ee
where the resulting variance $\sigma^2$ is given
by,
$$\sigma^2=\VAR\Big(
	\log\frac{W(Y|X)}{Q(Y)}
\Big)
=\sum_{x,y\in A}P(x)W(y|x)
	\Big(
	\log\frac{W(y|x)}{Q(y)}-I
	\Big)^2.$$
[An outline of the proof of (\ref{eq:clt}) 
is given below.]

Therefore, for any fixed threshold $\theta<I$
and large $n$,
the probability of error satisfies,
\begin{eqnarray}
P_{e,2}
&=&
        \PR\{\mbox{declare independence}\,|\,
        W\mbox{-dependent strings}\}
	\nonumber\\
&=&
        \PR\{I(n)\leq \theta\,|\,
        W\mbox{-dependent strings}\}
	\nonumber\\
&\approx&
        \PR\{V\leq [\theta-I]\sqrt{n}\}
	\nonumber\\
&\approx&
	\exp\Big\{-\frac{(I-\theta)^2}{2\sigma^2}\,n\Big\},
	\label{eq:p2}
\end{eqnarray}
where the last approximation sign indicates
equality to first order in the exponent.
Thus, despite the fact that $I(n)$ converges
at different speeds in the two scenarios,
both error probabilities $P_{e,1}$ and $P_{e,2}$ 
decay exponentially
with the sample size $n$.

To see why (\ref{eq:clt}) holds it is convenient
to use the alternative expression for $I(n)$ given 
in (\ref{eq:alt}).
Using this, and recalling that 
$I(n)=\hat{I}_1(n)$,
we obtain,
$$\sqrt{n}[I(n)-I]=\sqrt{n}
	\Big[
        \frac{1}{n}\sum_{i=1}^n\log\frac{\hat{p}_1(X_i,Y_i)}
        {\hat{p}(X_i)\hat{q}_1(Y_i)}
	-I\Big].
$$
Since the empirical distributions converge to the 
corresponding true distributions, for large $n$
it is straightforward to justify the approximation,
\be
	\sqrt{n}[I(n)-I]\approx
        \frac{1}{\sqrt{n}} 
	\Big[ 
	\sum_{i=1}^n\log\frac{P(X_i)W(Y_i|X_i)}
        {P(X_i)Q(Y_i)}
	-I\Big].
	\label{eq:approx}
\ee
The fact that this indeed 
converges in distribution to a
$N(0,\sigma^2)$, as $n\to\infty$,
easily follows from the central limit theorem,
upon noting that the mean of the
logarithm in (\ref{eq:approx}) equals $I$
and its variance is $\sigma^2$.

\paragraph{Discussion. } From the above analysis 
it follows that, in order for both probabilities of
error to decay to zero for large $n$ (so that we rule
out false positives as well as making sure that no dependent
segments are overlooked) the threshold $\theta$ needs to be
strictly between 0 and $I=I(X;Y)$. For that, we need to
have some prior information about the value of $I$,
i.e., of the level of dependence we are looking for.
If the value of $I$ were actually known
and a fixed threshold $\theta\in(0,I)$
was chosen independent of $n$, then
both probabilities of error would
decay exponentially fast, but
with typically very different exponents,
$$P_{e,1}\approx \exp\{-(\theta\ln 2)\, n\}
\;\;
\mbox{and}
\;\;
P_{e,2}\approx
\exp\Big\{-\Big(\frac{I-\theta}{\sqrt{2}\sigma}\Big)^2 n\Big\};$$
recall the expressions in (\ref{eq:p1}) and (\ref{eq:p2}).
Clearly, balancing
the two exponents also requires
knowledge of the value of $\sigma^2$ in
the case when the two strings are dependent,
which, in turn, requires full knowledge of the
marginal distribution $P$ and the channel $W$.
Of course this is unreasonable, since we cannot
specify in advance the exact kind and level
of dependence we are actually trying to detect
in the data.

A practical (and standard) approach 
is as follows:
Since the probability of error of the first kind
$P_{1,e}$ only depends on $\theta$
(at least for large $n$), and since
in practice declaring false positives
is much more undesirable than overlooking
potential dependence, in our experiments
we decide on an acceptably small false-positive
probability $\epsilon$, and then select $\theta$
based on the above approximation, by setting
$P_{e,1}\approx \epsilon$
in (\ref{eq:false_positives}).

\section{Experimental Results}
\label{experiments}

In this section we apply the mutual
information test described above to
biological data. First we show that 
it can be used effectively to identify 
statistical dependence between regions 
of the maize zmSRp32 gene that may be involved
in alternative processing (splicing) 
of pre-mRNA transcripts. 
Then we show how the same methodology
can be easily adapted to the problem 
of identifying tandem repeats.
We present experimental results 
on DNA sequences from the FBI's Combined 
DNA Index System (CODIS), which clearly
indicate that the empirical mutual 
information can be a powerful tool 
for this computationally intensive task.

\subsection{Detecting DNA Sequence Dependencies}
\label{DNASD}

\begin{figure}
\centerline{
\epsfig{file=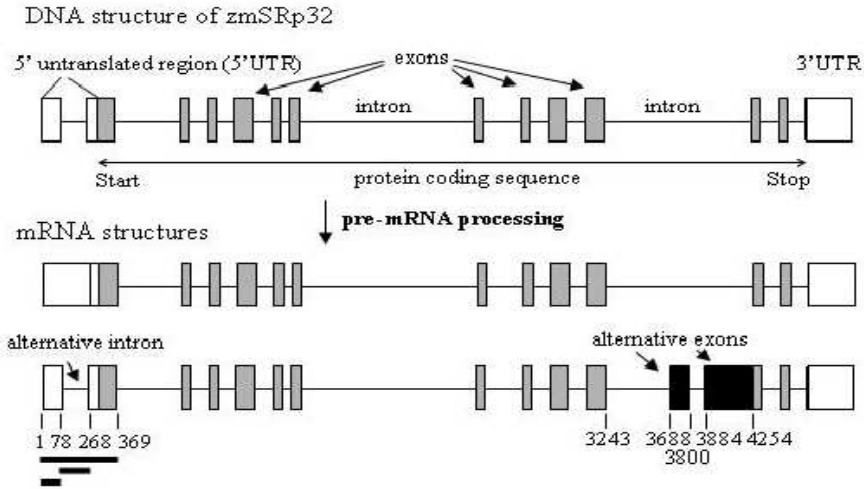, height=2.8in, width=5in}}
\caption{ 
Alternative splicings of the zmSRp32 gene in maize. The gene
consists of a number of exons (shaded boxes) and introns (lines) flanked
by the 5' and 3' untranslated regions (white boxes). RNA transcripts
(pre-mRNA) are processed to yield mRNA molecules used as templates for
protein synthesis. Alternative pre-mRNA splicing generates different
mRNA templates from the same transcripts, by selecting either alternative
exons or alternative introns. The regions discussed in the text are
identified by indices corresponding to the nucleotide position in the
original DNA sequence.}
\label{fig:zmSRp32}
\end{figure}

All of our experiments were performed 
on the maize 
zmSRp32 gene \cite{lyznik04}. 
This gene belongs to  a group of genes that are
functionally homologous to the human ASF/SF2 alternative 
splicing factor. Interestingly, these genes
encode alternative splicing factors in maize 
and yet themselves are also alternatively spliced. 
The gene zmSRp32 is coded by $4735$ nucleotides and has 
four alternative splicing variants. 
Two of these four variants are due to 
different splicings of this gene, between 
positions 1--369 and 3243--4220, respectively,
as shown in Figure~\ref{fig:zmSRp32}.
The results given here are primarily from experiments 
on these segments of zmSRp32.

\begin{figure}
\centerline{
\epsfig{file = 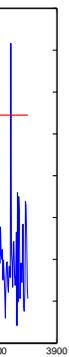,width=\basswidth}\hspace*{\gutter}
\epsfig{file = 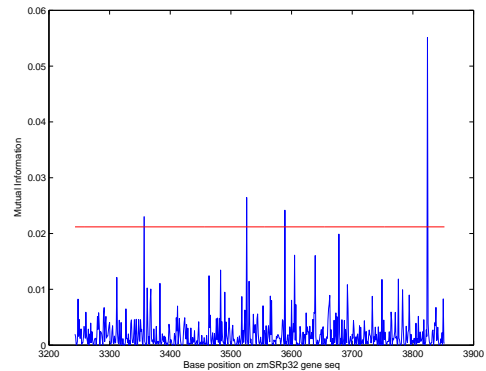,width=\basswidth}
}
\medskip
\centerline{\hbox to \basswidth{\hfill(a)\hfill}\hspace*{\gutter}
        \hbox to \basswidth{\hfill(b)\hfill}}
\caption{ 
Estimated mutual information between the exon 
located between bases 1--369 and each 
contiguous subsequence of length 369 
in the intron between bases 3243--4220.
The estimates were computed both for the
original sequences in the standard four-letter 
alphabet $\{A,C,G,T\}$ (shown in (a)), 
as well as for the corresponding transformed sequences 
for the two-letter purine/pyrimidine grouping 
$\{AG, CT\}$ (shown in (b)).
\normalsize
}
\label{mutinf:good1}
\end{figure}

In order to understand and quantify the amount of 
correlation between different parts of this gene, 
we computed the mutual information between 
all functional elements including exons, introns, 
and the 5' untranslated region. As before, we denote 
the shorter sequence of length $n$ by $X_1^n=(X_1,X_2,\ldots,X_n)$ 
and the longer one of length $M$ by $Y_1^M=(Y_1,Y_2,\ldots,Y_M)$. 
We apply the 
simple mutual information estimator $\hat{I}_j(n)$ 
defined in (\ref{eq:hat{I}_j}) to estimate the mutual 
information between $X_1^n$ and $Y_j^{j+n-1}$
for each $j=1,2,\ldots,M-n+1$, and we plot
the ``dependency graph'' of
$\hat{I}_j=\hat{I}_j(n)$ versus $j$; 
see Figure~\ref{mutinf:good1}.
The threshold $\theta$ is computed according to
(\ref{eq:false_positives}), by setting 
$\epsilon$, the probability of false positives, 
equal to $0.001$; it is represented by a (red) 
straight horizontal line in the figures.

\begin{figure}
\centerline{
  \epsfig{file = 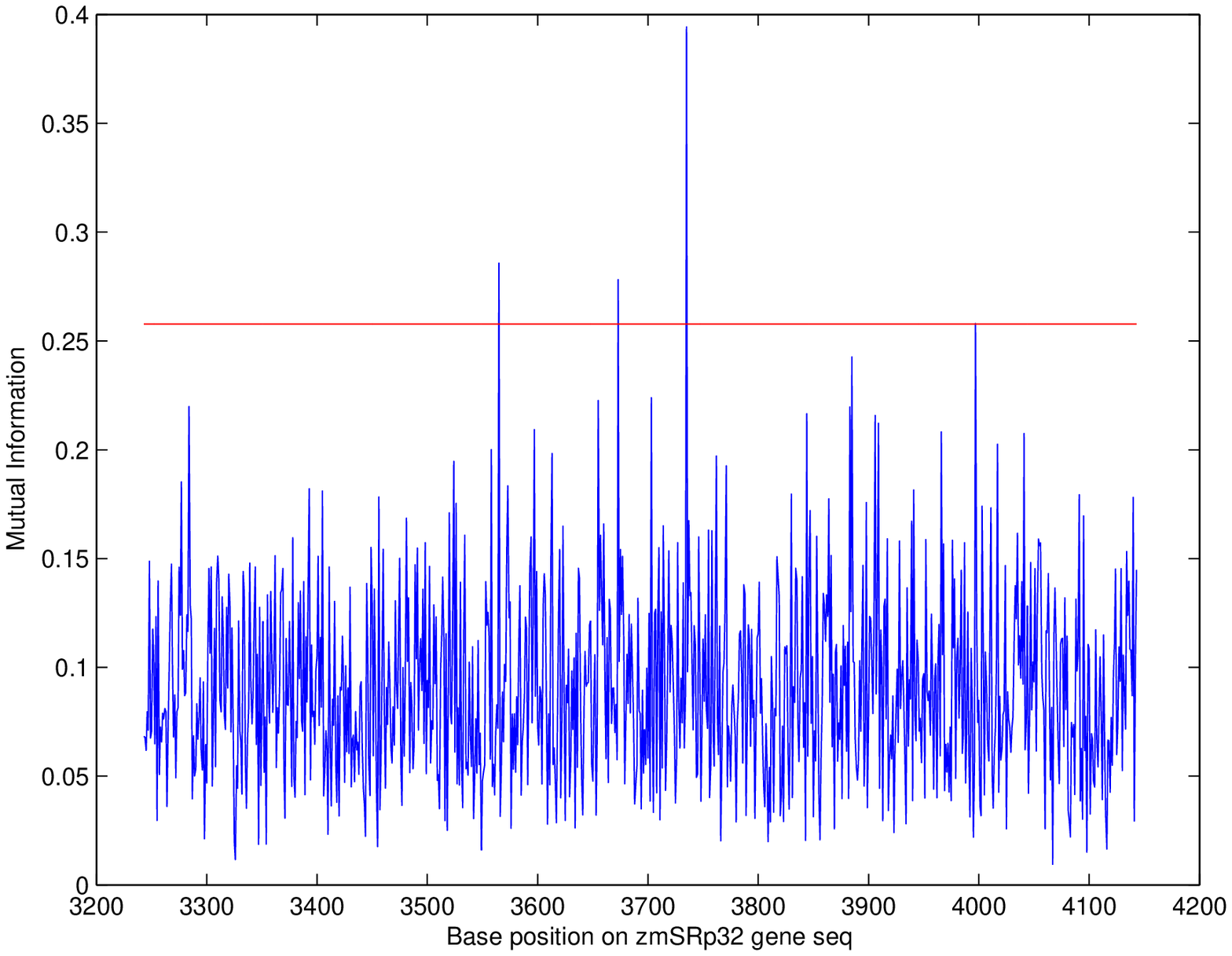,width=\basswidth}\hspace*{\gutter}
  \epsfig{file = 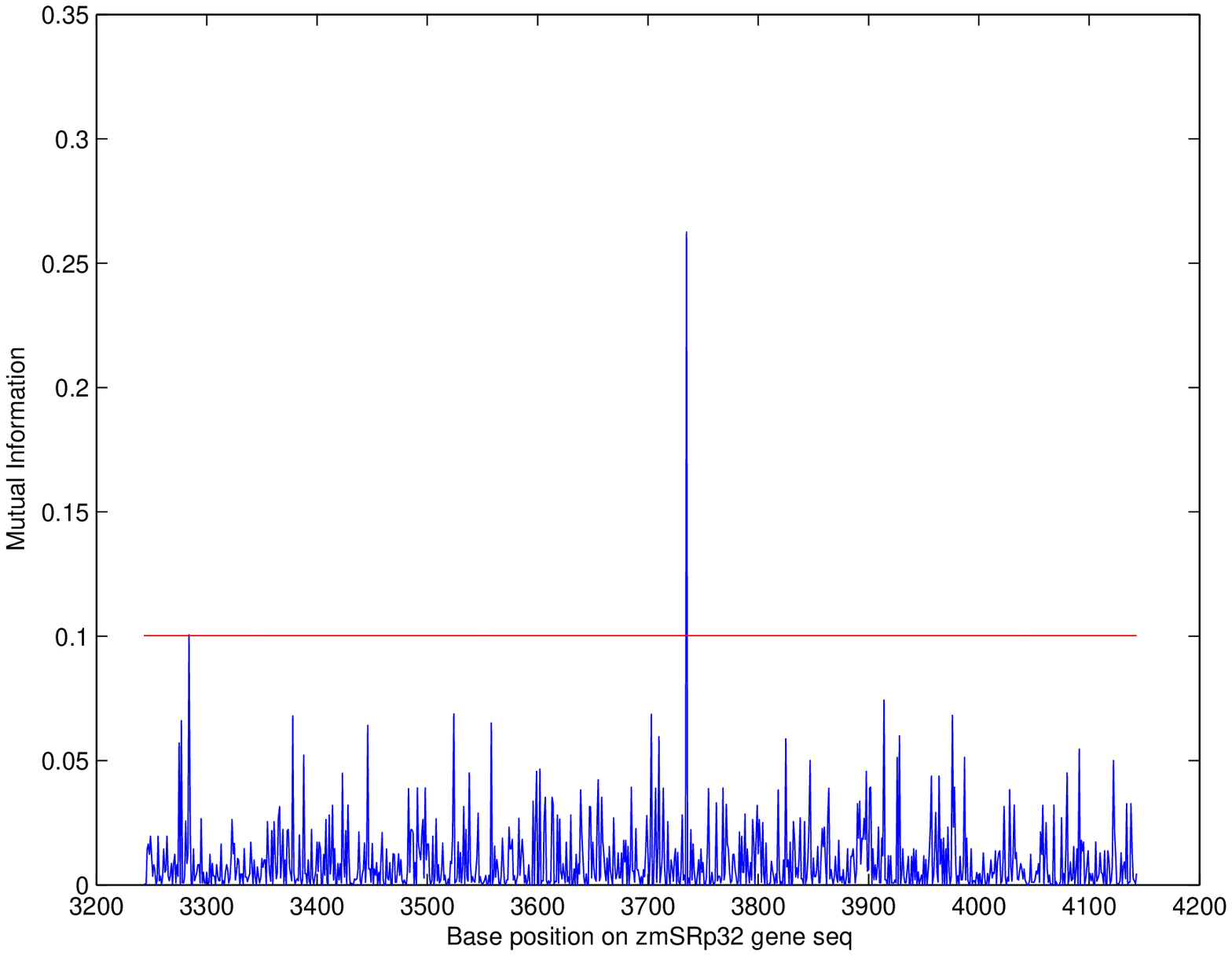,width=\basswidth}
}
\medskip
\centerline{\hbox to \basswidth{\hfill(a)\hfill}\hspace*{\gutter}
  \hbox to \basswidth{\hfill(b)\hfill}
}

\centerline{
  \epsfig{file = 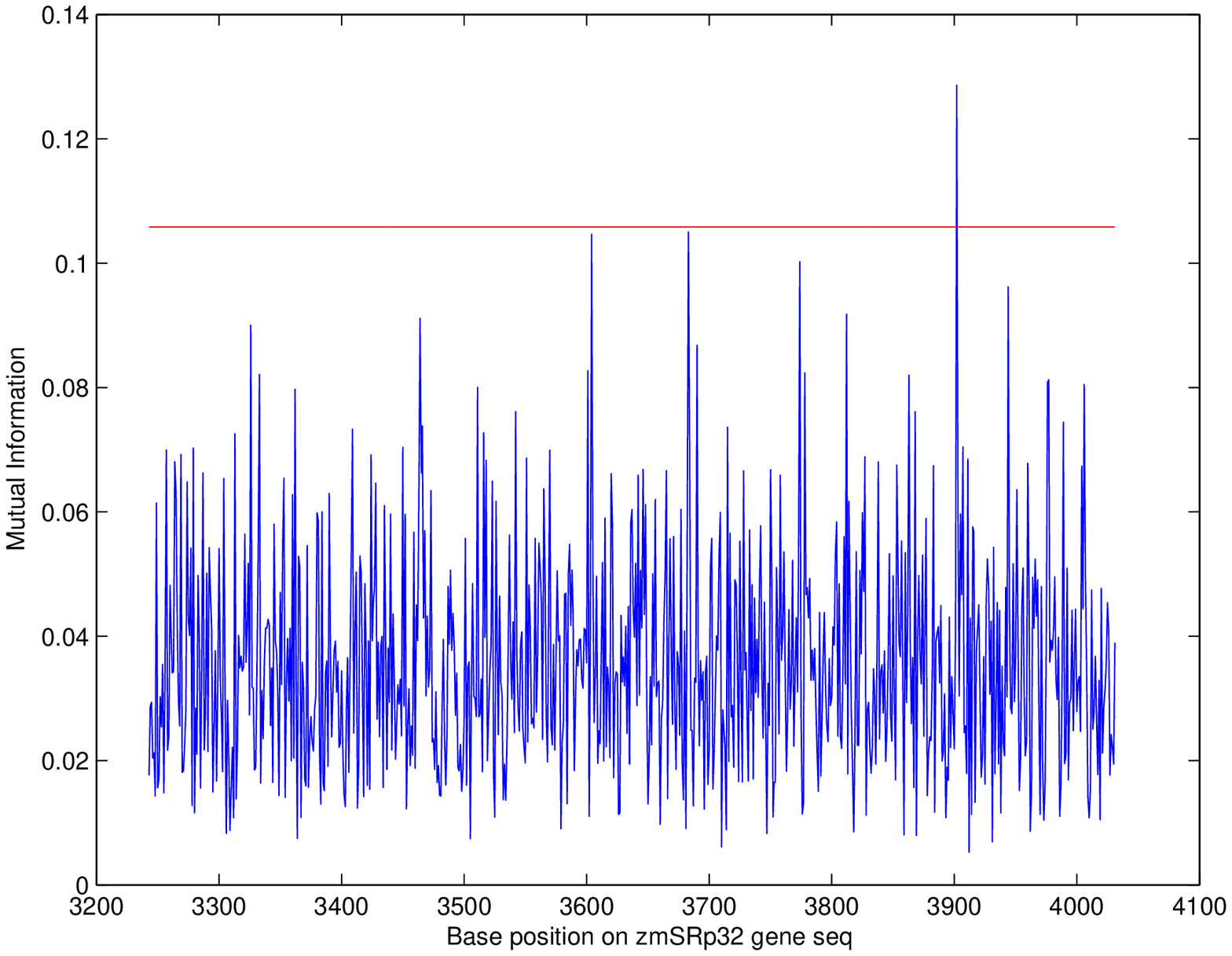,width=\basswidth}\hspace*{\gutter}
  \epsfig{file = 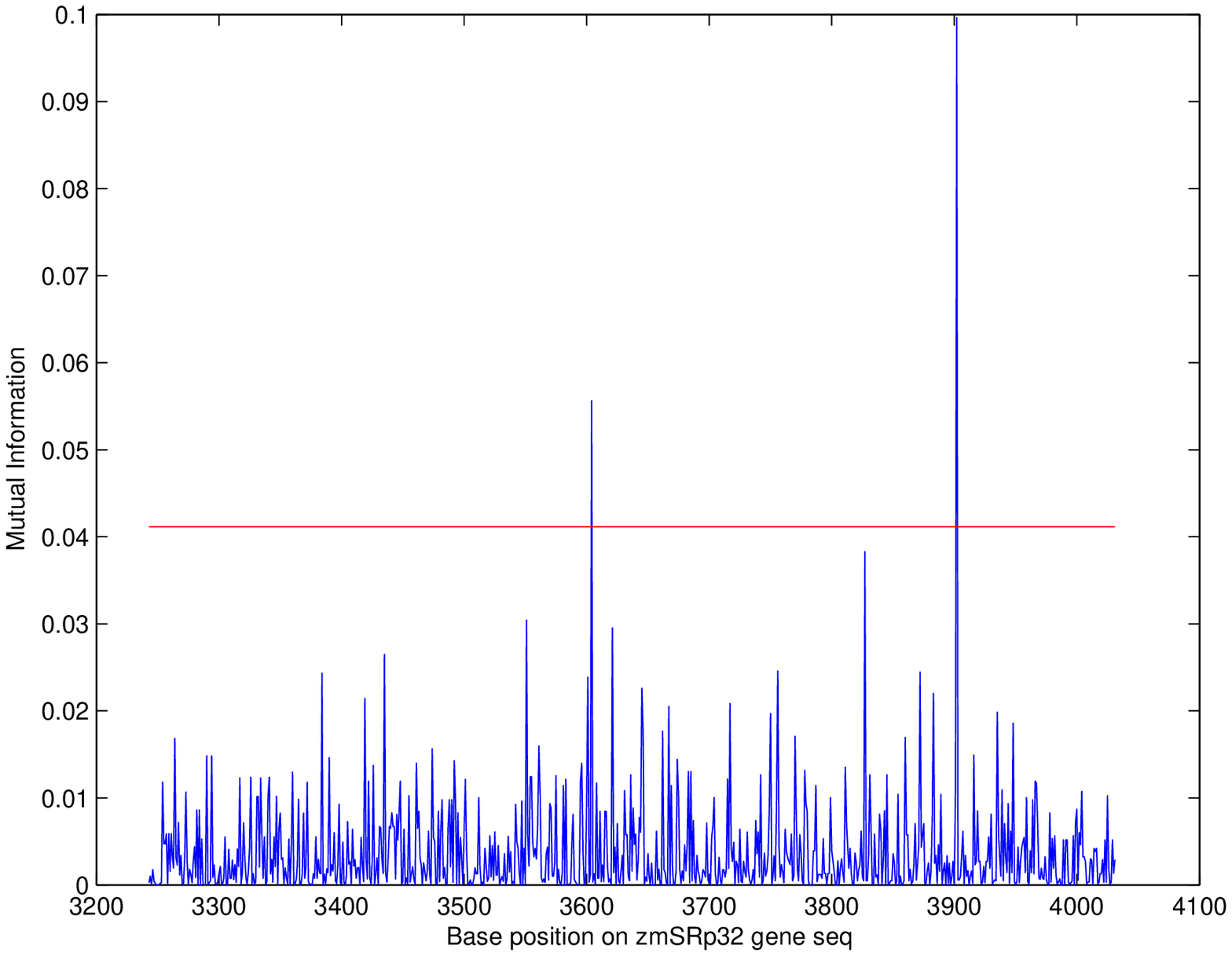,width=\basswidth}
}
\medskip
\centerline{\hbox to \basswidth{\hfill(c)\hfill}\hspace*{\gutter}
  \hbox to \basswidth{\hfill(d)\hfill}
}

\centerline{
  \epsfig{file = 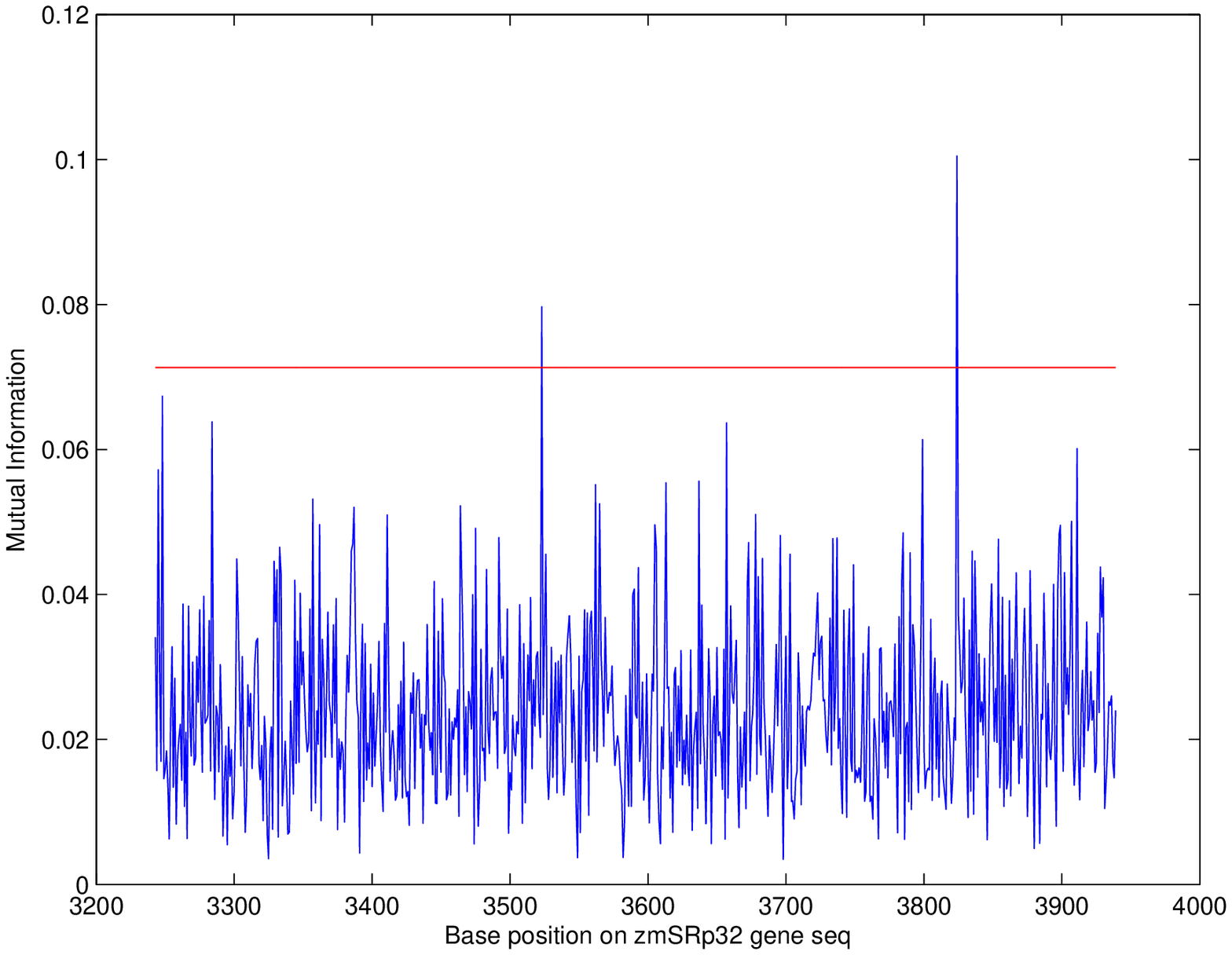,width=\basswidth}\hspace*{\gutter}
  \epsfig{file = 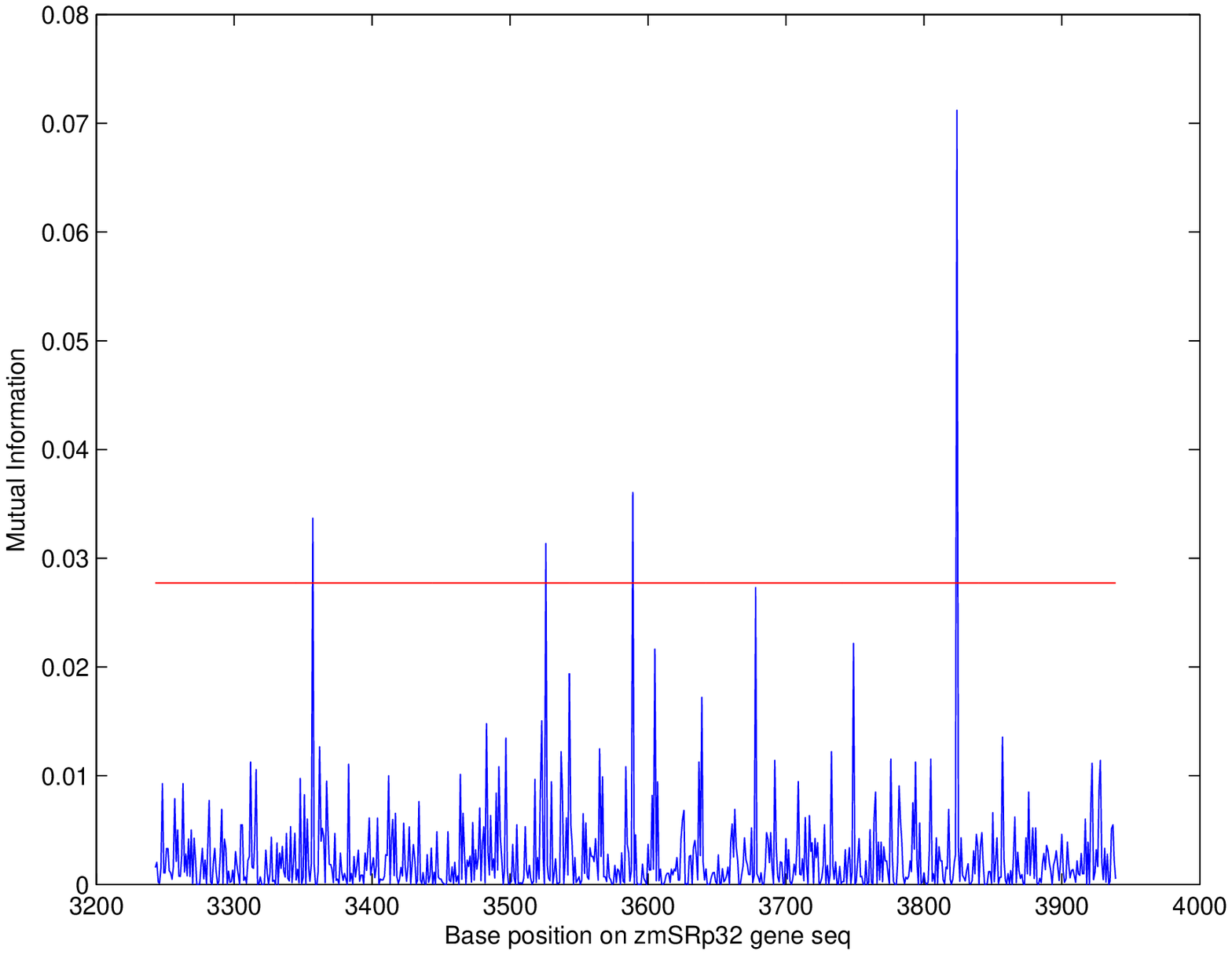,width=\basswidth}
}
\medskip
\centerline{\hbox to \basswidth{\hfill(e)\hfill}\hspace*{\gutter}
  \hbox to \basswidth{\hfill(f)\hfill}
}
\caption{
Dependency graph of $\hat{I}_j$ versus $j$ for the zmSRp32 
gene, using different alphabet groupings:
In (a) and (b), we plot the estimated mutual information 
between the exon found between bases 1--78 and 
each subsequence of length 78 in the intron 
located between bases 3243--4220. Plot~(a) shows
estimates over the original
four-letter alphabet $\{A,C, G,T\}$, and~(b) shows
the corresponding estimates 
over the Watson-Crick pairs $\{AT,CG\}$.
Similarly, plots~(c) and~(d) contain the
estimated mutual information between the intron located 
in bases 79--268 and all corresponding subsequences 
of the intron between bases 3243--4220. 
Plot~(c) shows estimates over the
original alphabet,  and plot~(d) over 
the two-letter purine/pyrimidine grouping  $\{AG, CT\}$.
Plots~(e) and~(f) show the estimated mutual information 
between the 5' untranslated region 
and all corresponding subsequences of the intron 
between bases 3243--4220, for the 
four-letter alphabet (in~(e)), and for the two-letter 
purine/pyrimidine grouping $\{AG, CT\}$ (in~(f)).
}
\label{mutinf:good2}
\end{figure}

In order to ``amplify'' the effects of 
regions of potential dependency in
various segments of the zmSRp32 gene,
we computed the mutual information 
estimates $\hat{I}_j$ 
on the original strings
over the regular four-letter alphabet $\{A,C,G,T\}$,
as well as on transformed versions of the strings
where pairs of letters were grouped together,
using either the Watson-Crick pair $\{AT,CG\}$ or 
the purine-pyrimidine pair $\{AG,CT\}$. 
In our results we observed that such groupings
are often helpful in 
identifying dependency; this is clearly 
illustrated by the estimates shown
in Figures~\ref{mutinf:good1} and~\ref{mutinf:good2}. 
Sometimes the $\{AT,CG\}$ pair produces better
results, while in other cases the purine-pyrimidine pair 
finds new dependencies.

Figure~\ref{mutinf:good1} strongly suggests
that there is significant dependence between 
the bases in positions 1--369 and certain 
substrings of the bases in positions 3243--4220.
While the 1--369 region contains the 5' untranslated sequences, 
an intron, and the first protein coding exon, the 3243--4220 sequence
encodes an intron that undergoes alternative splicing. 
After narrowing down the mutual information calculations 
to the 5' untranslated region (5'UTR) in positions 1--78 
and the 5'UTR intron in positions 78--268,
we found that the initially identified dependency 
was still present; see Figure~\ref{mutinf:good2}.
A close inspection of the resulting mutual information
graphs indicates 
that the dependency is restricted to the alternative 
exons embedded into the intron sequences,
in positions 3688--3800 and 3884--4254. 

These findings suggest that there might be 
a deeper connection between 
the 5'UTR DNA sequences and the DNA sequences that 
undergo alternative splicing. The UTRs are multifunctional genetic 
elements that control gene expression by determining mRNA stability 
and efficiency of mRNA translation. Like in the zmSRp32 maize gene, 
they can provide multiple alternatively spliced variants for more complex 
regulation of mRNA translation \cite{hughes06}. They also contain a number 
of regulatory motifs that may affect many aspects of mRNA metabolism. 
Our observations can therefore be interpreted as
suggesting that the maize zmSRp32 5'UTR contains information
that could be utilized in the process of alternative 
splicing, yet another important aspect of mRNA metabolism. 
The fact that the value of the empirical mutual information 
between 5'UTR and the DNA sequences that encode 
alternatively spliced elements is significantly
greater than zero clearly points 
in that direction. Further experimental work
could be carried out to verify the existence,
and further explore the meaning, of these 
newly identified statistical dependencies.

We should note that there
are many other sequence matching 
techniques, the most popular of which 
is probably the celebrated BLAST algorithm.
BLAST's working principles are very different 
from those underlying our method. As a first step, BLAST 
searches a database of biological sequences 
for various small words found in the query string. 
It identifies sequences that are 
candidates for potential matches, and thus
eliminates a huge portion of the database containing sequences 
unrelated to the query.  In the second step, small word matches 
in every candidate sequence are extended by means 
of a Smith-Waterman-type local 
alignment algorithm. Finally, these extended local alignments
are combined with some scoring schemes, and the highest 
scoring alignments obtained are returned. 
Therefore,
BLAST requires a considerable fraction of exact matches to 
find sequences related to each other. However,
our approach does not enforce any such requirements. 
For example, if two sequences do not have any exact matches at all, 
but the characters in one sequence are a character-wise
encoding of the ones in the other sequence,
then BLAST would fail to produce any significant 
matches (without corresponding substitution matrices), 
while our algorithm would detect a 
high degree of dependency. This is illustrated
by the results in the following section, where 
the presence of certain repetitive patterns in $Y_1^M$
is revealed through matching it to a ``probe sequence''
$X_1^n$ which does {\em not} contain the repetitive
pattern, but is ``statistically similar'' to the 
pattern sought.

\subsection{Application to Tandem Repeats}
\label{STRs}
Here we further explore the utility of the mutual
information statistic, and we examine its performance
on the problem of detecting Short Tandem Repeats (STRs) 
in genomic sequences. 
STRs, usually found in non-coding regions, 
are made of back-to-back repetitions of a sequence 
which is at least two bases long and generally shorter
than $15$ bases.
The period of an STR is defined as the length of
the repetition sequence in it.
Owing to their short lengths, STRs survive mutations well, and 
can easily be amplified using PCR without producing erroneous data. 
Although there are many well-identified STRs in the human genome,
interestingly, the number of repetitions 
at any specific locus varies significantly 
among individuals; that is, 
they are {\it polymorphic} DNA fragments. These properties 
make STRs suitable tools for determining genetic profiles, and 
have become a prevalent method in forensic investigations. 
Long repetitive sequences have also been observed in genomic sequences,
but have not gained as much attention since they cannot survive 
environmental degradation and do not produce high quality data from
PCR analysis. 

Several algorithms have been proposed for detecting STRs in 
long DNA strings
with no prior knowledge about the size and the pattern of repetition.
These algorithms are mostly based on pattern matching,  and they
all have high time-complexity.
Finding short repetitions in a long sequence is a challenging problem.
When the query string is a DNA segment that contains many 
insertions, deletions
or substitutions due to mutations, the problem becomes even harder.
Exact- and approximate-pattern matching algorithms need to be modified to 
account for these mutations, and this 
renders them complex and inefficient.
To overcome these limitations,
we propose a statistical approach using an adaptation
of the method described in the previous sections.

In the United States, the FBI has decided on $13$ loci to be used as the 
basis for genetic profile analysis, and they continue to be 
the standard in this area. To demonstrate how our approach can be used 
for STR detection, we chose to use 
sequences from the FBI's Combined DNA Index System (CODIS):
The SE33 locus contained in the GenBank sequence V00481, 
and the VWA locus contained in the GenBank sequence M25858.
The periods of STRs found in CODIS typically range from 
2 to 4 bases, and do not exhibit enough variability 
to demonstrate how our approach would perform under 
divergent conditions. For this reason, we used
the V00481 sequence as is, but on M25858 we artificially 
introduced an STR with period 11, by substituting bases 2821--2920 
(where we know that there are no other repeating sequences)
with $9$ tandem repeats of $ACTTTGCCTAT$.
We have also introduced base substitutions, deletions 
and insertions on our artificial STR to imitate mutations.

\begin{figure}
\centerline{
\epsfig{file=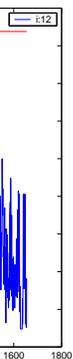,width=\basswidth}\hspace*{\gutter}
\epsfig{file=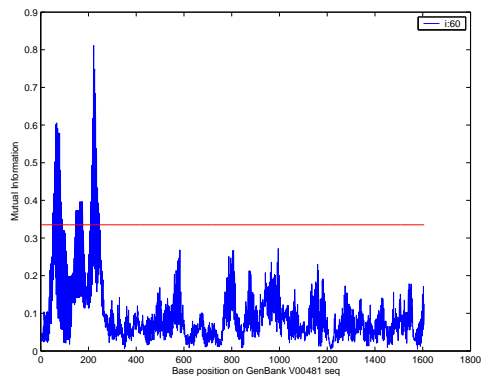,width=\basswidth}}
\medskip
\centerline{\hbox to \basswidth{\hfill(a)\hfill}\hspace*{\gutter}
        \hbox to \basswidth{\hfill(b)\hfill}}
\caption{
Dependency graph of the GenBank sequence $Y_1^M=V00481$, for 
a probe sequence $X_1^n$ which is 
a repetition of $AGGT$, of length: (a)~$12$, or (b)~$60$.
The sequence $Y_1^M$ contains STRs that are 
repetitions of the pattern $AAAG$,
in the following regions: (i) there is a 
repetition of $AAAG$ between bases 
62--108; (ii) $AAAG$ is intervened by $AG$ and $AAGG$ until 
base $138$; (iii) again between 138--294 there are 
repetitions of $AAAG$, some of which are modified by 
insertions and substitutions. 
In~(a) our probe is too short, and it is almost impossible 
to distinguish the SE33 locus from the rest. However, in~(b) 
the location SE33 is singled out by the two
big peaks in the mutual information estimates; 
the shorter peak between the two larger ones 
is due to the interventions described above.
Note that the STRs were identified by a probe 
sequence that was a repetition of a pattern
{\em different} from that of the repeating part of 
the STRs themselves,
but of the same period.}
\label{fig:Rlength}
\end{figure}

Let $Y_1^M=(Y_1,Y_2,\ldots,Y_M)$ denote the DNA sequence 
in which we are looking for STRs.
The gist of our approach is simply to choose 
a {\em periodic} probe sequence of length $n$,
say, $X_1^n=(X_1,X_2,\ldots,X_n)$ (typically
much shorter than $Y_1^M$), and then to 
calculate the empirical mutual information 
$\hat{I}_j=\hat{I}_j(n)$
between $X_1^n$ and each of its possible 
alignments with $Y_1^M$.
In order to detect the presence of STRs, 
the values of the empirical mutual information 
in regions where STRs do appear should be 
significantly larger than zero, where
``significantly'' means larger than the
corresponding estimates in
ordinary DNA fragments containing no STRs.
Obviously, the results will depend heavily
on the exact form of the probe sequence.
Therefore, it is critical to decide on the
method for selecting: (a)~the length, and 
(b)~the exact contents of $X_1^n$.
The length of $X_1^n$ is crucial; if it is too short, 
then $X_1^n$ itself is likely to appear often in
$Y_1^M$, producing many large values of the empirical 
mutual information and making it hard to distinguish 
between STRs and ordinary sequences. Moreover, in that
case there is little hope that the analysis
of the previous section (which was carried out
of long sequences $X_1^n$) will provide useful estimates
for the probability of error.
If, on the other hand, $X_1^n$ is too long,
then any alignment of the probe $X_1^n$ with
$Y_1^M$ will likely also 
contain too many irrelevant base pairs.
This will produce negligibly small mutual
information estimates, again making impossible
to detect STRs.
These considerations 
are illustrated by the results
in Figure~\ref{fig:Rlength}.

\begin{figure}
\centerline{
\epsfig{file=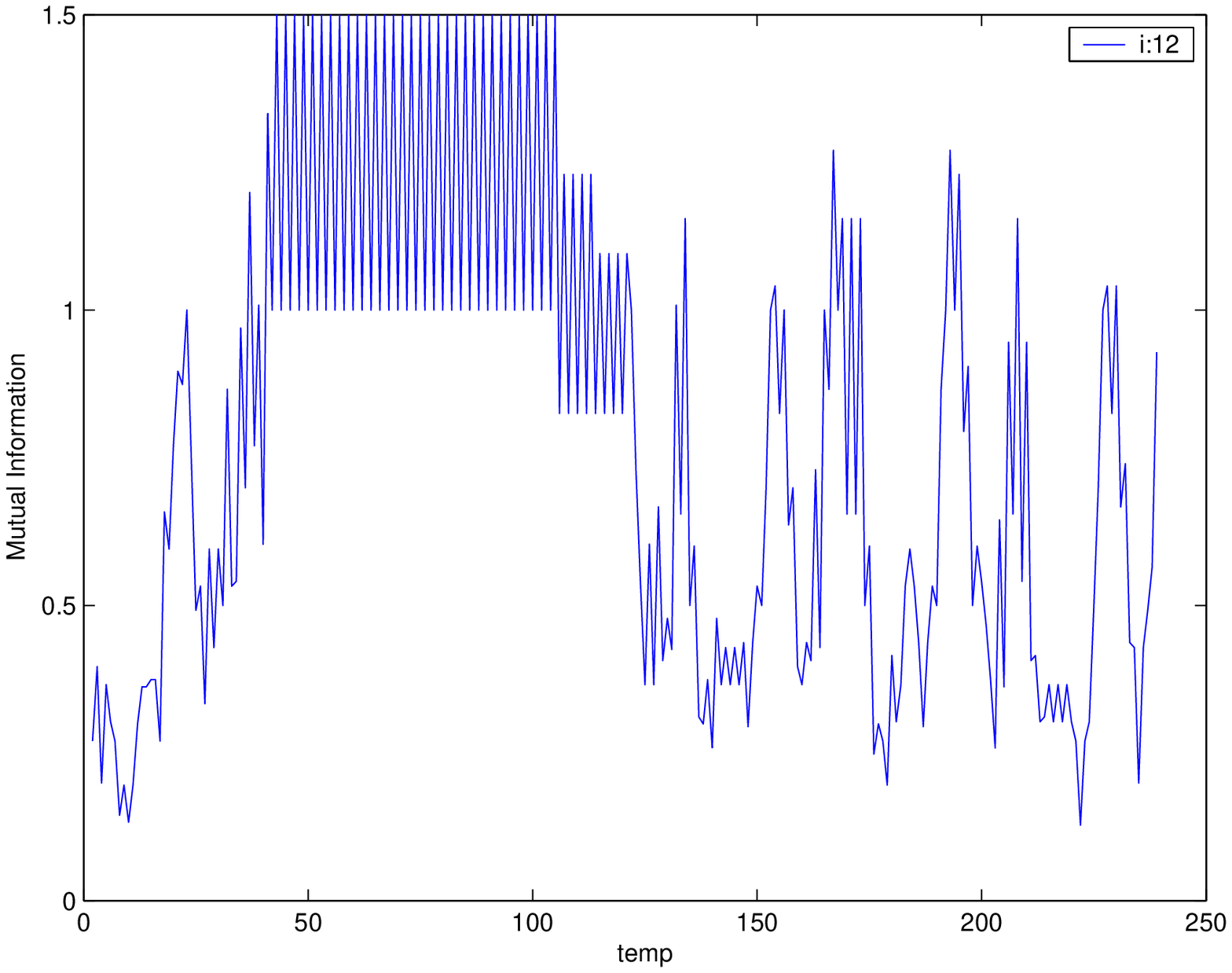,width=\basswidth}\hspace*{\gutter}
\epsfig{file=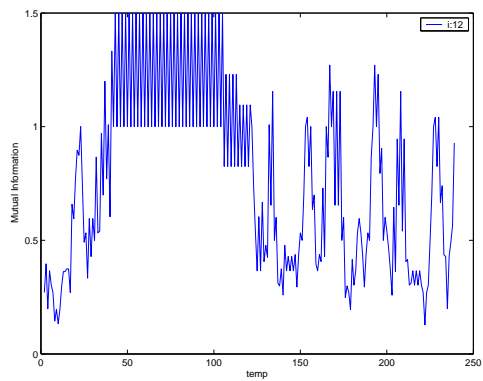,width=\basswidth}}
\medskip
\centerline{\hbox to \basswidth{\hfill(a)\hfill}\hspace*{\gutter}
        \hbox to \basswidth{\hfill(b)\hfill}}
\caption{
Dependency graph of the VWA locus contained in GenBank sequence M25858 
for a probe sequence $X_1^n$ with $n=12$, which is a repetition of:
(a) $TCTA$, an exactly matching probe,
(b) $GTGC$, a completely different probe,
but of the exact same ``pattern.'' 
In both cases, 
we have chosen $X_1^n$ to be long enough
to suppress unrelated information.
Note that the results in (a) and (b) are almost
identical.
The VWA locus contains an STR of $TCTA$ between positions 44--123. 
This STR is apparent in both dependency graphs by 
forming a periodic curve with high correlation.
}
\label{fig:STR_exact_sim}
\end{figure}

As for the contents
of the probe sequence $X_1^n$, 
the best choice would be to take
a segment $X_1^n$ containing an exact match 
to an STR present in $Y_1^M$.
But in most of the interesting applications,
this is of course unavailable to us.
A ``second best'' choice might be a 
sequence $X_1^n$ that contains a segment 
of the same ``pattern'' as the STR present in $Y_1^M$,
where we say that two sequences have the same {\em pattern}
if each one can be obtained from the other via a 
permutation of the letters in the alphabet;
cf.\ \cite{shtarkov,patterns}.
For example, $TCTA$ and $GTGC$ have the same pattern,
whereas $TCTA$ and $CTAT$ do not (although the do
have the same empirical distribution).
For example, if $X_1^n$ contains the exact same pattern 
as the periodic part of the STR to be detected, 
and $\tilde{X}_1^n$ has the same pattern as $X_1^n$,
then, a priori, either choice should be equally
effective at
detecting the STR under consideration;
see Figure~\ref{fig:STR_exact_sim}.
[This observation also shows that a single probe $X_1^n$
may in fact be appropriate for locating more
than a single STR; for example, STRs with the
same pattern as 
$X_1^n$, as in Figure~\ref{fig:STR_exact_sim},
or with the same period, as in Figure~\ref{fig:Rlength}.]
The problem with this choice is, again,
that the exact patterns 
of STRs present in a DNA sequence 
are not available to us in advance, and we cannot expect 
all STRs in a given sequence to be of the same 
pattern.

Even though both of the above choices for $X_1^n$
are usually not practically feasible,
if the sequence $Y_1^M$ is relatively 
short and contains a single STR whose contents are known, 
then either choice would produce high 
quality data, from which the STR contained in $Y_1^M$ 
we can easily be detected; see 
Figure~\ref{fig:STR_exact_sim}
for an illustration.

\begin{figure}
\centerline{
\epsfig{file=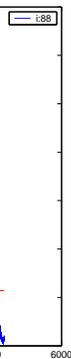,width=\basswidth}\hspace*{\gutter}
\epsfig{file=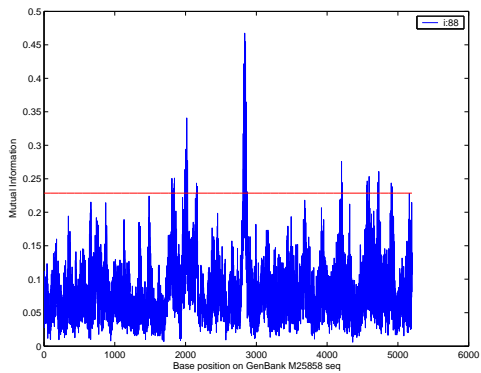,width=\basswidth}}
\medskip
\centerline{\hbox to \basswidth{\hfill(a)\hfill}\hspace*{\gutter}
        \hbox to \basswidth{\hfill(b)\hfill}}
\caption{
In these charts we use the modified GenBank sequence M25858,
which contains the VWA locus in CODIS between positions 1683--1762 and
the artificial STR introduced by us at 2821--2920.
The repeat sequence of the VWA locus is $TCTA$, 
and the repeat sequence of the artificial STR is
$ACTTTGCCTAT$.
In (a), the probe $X_1^n$ has length $n=88$ and consists
of repetitions of $AGGT$.
Here the repeating sequence of the VWA locus 
(which has period 4) is clearly indicated by the peak, 
whereas the artificial tandem repeat (which has period 11)
does not show up in the results.
The small 
peak around position 2100 is due to a very noisy STR again with a 4 base period.
In (b), the probe $X_1^n$ again has length $n=88$, and it 
consists of repetitions of $CATAGTTCGGA$. This produces
the opposite result: The artificial STR is clearly identified,
but there is no indication of the STR present at 
the VWA locus.}
\label{fig:Rperiod}
\end{figure}

In practice, in addition to the fact that
the contents of STRs are not known in advance,
there is also the issue that in a long DNA sequence 
there are often many different STRs, and a unique 
probe will not match all of them exactly.
But since STRs usually have a period between 
2 and 15 bases, we can actually run our method 
for all possible choices of repetition sequences,
and detect all STRs in the given query sequence 
$Y_1^M$. The number of possible probes $X_1^n$ 
can be drastically by observing that 
(1)~We only need one repeating sequence of each
possible pattern; 
and (2)~It suffices to only consider repetition patters
whose period is prime.
Note that, in view of the earlier discussion
and the results shown in Figure~\ref{fig:Rlength},
the period of the repeating part of $X_1^n$ is 
likely to be more important than the actual contents.
For example, if we were to apply our method for finding 
STRs in $Y_1^M$ with a probe $X_1^n$ whose period is $5$ bases
long, then many STRs with a period that is a multiple of $5$ 
should peak in the dependency chart, thus allowing us to detect 
their approximate positions in $Y_1^M$. 
Clearly, probes that consist of very short 
repeats, such as $AAA\ldots...$, should 
be avoided.
The importance of choosing an $X_1^n$ with the correct 
period is illustrated in Figure~\ref{fig:Rperiod}.

The results in Figures~\ref{fig:Rlength},
\ref{fig:STR_exact_sim}
and~\ref{fig:Rperiod} clearly indicate
that the proposed methodology is very
effective at detecting the presence of
STRs, although at first glance it may appear 
that it cannot provide precise information 
about their start-end positions and their
repeat sequences.
But this final task can easily be accomplished
by re-evaluating $Y_1^M$ near the peak 
in the dependency graph, for example,
by feeding the relevant parts separately into 
one of the standard
string matching-based tandem repeat algorithms.
Thus, our method can serve as an initial filtering 
step which, combined with an exact pattern matching algorithm,
provides a very accurate and efficient method for the
identification of STRs.

In terms of its practical implementation,
note that our approach has a linear running time $O(M)$, 
where $M$ is the length of $Y_1^M$. 
The empirical mutual information 
of course needs to be evaluated
for every possible alignment of $Y_1^M$ 
and $X_1^n$, with each such calculation 
done in $O(n)$ steps, where $n$
is the length of $X_1^n$. 
But $n$ is typically no longer than a few hundred bases, 
and, at least to first order, it can be considered constant.
Also, repeating this process for all possible repeat periods 
does not affect the complexity of our method by much, 
since the number of such periods is quite small and can 
also be considered to be constant. And, as mentioned above,
choosing probes $X_1^n$ only containing repeating segments 
with a prime period, further improves 
the running time of our method.

We, therefore, conclude that:
(a)~the empirical mutual information appears
in this case to be a very effective tool for
detecting STRs; and (b)~selecting the length
and repetition period of the probe sequence 
$X_1^n$ is crucial for identifying tandem 
repeats accurately.

\section{Conclusions}

Biological information is stored in the form of monomer strings
composed of conserved biomolecular sequences. 
According to Manfred Eigen, 
``The differentiable characteristic of living systems 
is information. Information assures the controlled
reproduction of all constituents, thereby ensuring 
conservation of viability.'' 
Hoping to reveal novel, potentially important
biological phenomena, we employ information-theoretic 
tools, especially the notion of mutual information,
to detect statistically dependent segments of biosequences.  
The biological implications of the existance of such correlations 
are deep, and they themselves remain unresolved.
The proposed approach may provide a powerful key
to fundamental advances in understanding and 
quantifying biological information. 

This work addresses two specific applications based on the
proposed tools. From the experimental analysis carried out 
on regions of the maize zmSRp32 gene, our findings suggest 
the existence of a biological connection between the 5' untranslated 
region in zmSRp32 and its alternatively spliced exons, 
potentially indicating the presence of novel alternative 
splicing mechanisms or structural scaffolds. 
Secondly, through extensive analysis of CODIS data, we show
that our approach is particularly well suited for the problem of
discovering short tandem repeats, an application of importance
in genetic profiling studies.


\end{document}